\documentclass{appolb}
\usepackage{graphicx}
\usepackage{amssymb}
\usepackage{lineno}
\usepackage{cite}

\begin{document}
\title{Femtoscopy of Proton, Light nuclei, and Strange hadrons in Au+Au Collisions at STAR
\thanks{Presented at Quark Matter 2022, Krakow, Poland, April 4-10, 2022}
}
\author{{Ke Mi (for the STAR Collaboration)
\address{Key Laboratory of Quark $\&$ Lepton Physics (MOE) and Institute of Particle Physics, Central China Normal University, Wuhan, 430079, China}
\address{Physics Institute, Heidelberg University, Heidelberg, 69117, Germany}
}}
\maketitle

\begin{abstract}
In these proceedings, we present the measurements of proton, light nuclei, and strange particle with neutral kaons correlation functions in Au+Au collisions at the BES program and top RHIC energy. The experimental results will be compared with model calculations to extract the size of emitting source and the properties of final state interactions. The collision energy and centrality dependence of the source size will be studied. Further, the implications for the production mechanism of light nuclei will be discussed.
\end{abstract}
  
\section{Introduction}
Measurements of the correlation function for a pair of particles with small relative momenta can provide insight into the geometry and lifetime of the particle-emitting source in relativistic heavy-ion collisions~\cite{Lisa:2005dd}. By measuring correlation at low relative velocities, one could access the smallest sizes in nature, which corresponds to the size of a nucleon (1 fm), therefore such a two-particle correlation method is called 'Femtoscopy'~\cite{Lednicky:2005af}.

The fundamental observable in femtoscopy is the correlation function $C(k^{*})$. The theoretical definition is expressed as a function of the relative distance between two particles $\mathbf{r}^{*}$ and their reduced relative momentum, $k^{*}$ = $\frac{1}{2}$ $*$ $|\mathbf{p}^{*}_{1} - \mathbf{p}^{*}_{2}|$ in the pair rest frame, with $\mathbf{p}^{*}_{1}$ = $-\mathbf{p}^{*}_{2}$, by the Koonin-Pratt formula~\cite{PhysRevD.33.1314}:
\begin{equation}
C\left(k^{*}\right)=\int S\left(\mathbf{r}^{*}\right)\left|\psi\left(\mathbf{r}^{*}, \mathbf{k}^{*}\right)\right|^{2} \mathrm{~d}^{3} r,
\end{equation}
where the first term $S(\mathbf{r}^{*})$ describes the source that emits the particles; the second term contains the interaction part via the two-particle wave function $\psi(\mathbf{r}^{*},\mathbf{k}^{*})$. 

The experimental correlation function is defined as a ratio of the probability of registering two particles simultaneously (in the same event) to the product of registering probabilities of such particles independently (in the mixed event):
\begin{equation}
C\left(k^{*}\right)= \frac{N_{\mathrm{same}}\left(k^{*}\right)}{N_{\mathrm{mixed}}\left(k^{*}\right)}.
\end{equation}
The shape and amplitude of the correlation function are determined by the quantum statistical effect (identical pair) and final state interaction (strong and Coulomb interactions).

\section{Particle Identification and Reconstruction}
\begin{figure}
\hspace{-1.8cm}
    \includegraphics[width=1.26\textwidth]{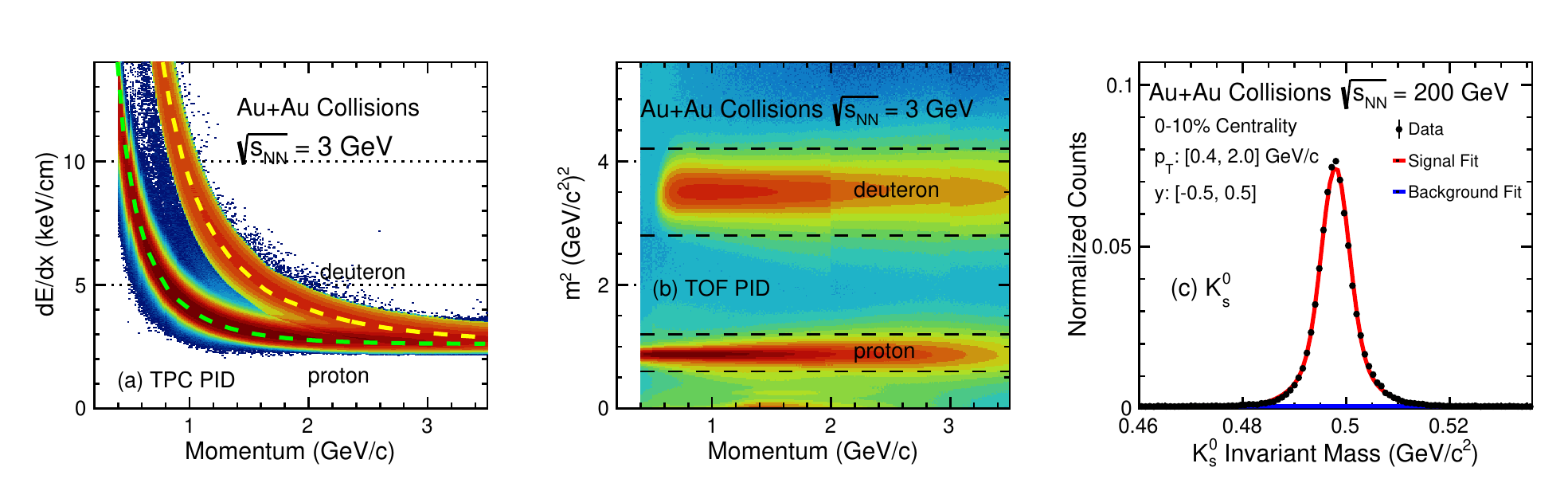}
    \caption{(a) The $dE/dx$ distribution of protons and deuterons versus total momentum in Au+Au collisions at $\sqrt{s_{\rm NN}}$ = 3\,GeV. The dashed curves are the corresponding Bichsel expectations. (b) Particle mass square ($m^{2}$) distribution versus total momentum in Au+Au collisions at $\sqrt{s_{\rm NN}}$ = 3\,GeV. The dashed lines represent the $m^{2}$ ranges used in the analysis. (c) The $K_{s}^{0}$ invariant mass distribution in 0-10\% centrality at $\sqrt{s_{\rm NN}}$ = 200\,GeV. The red line represents the fit to the data including signal and background, and blue line to the background alone.}
    \label{fig:PID_ks0}
\end{figure}

Particle identification (PID) of protons and deuterons are done by using the energy loss ($dE/dx$) information measured by Time Projection Chamber (TPC) (Fig.~\ref{fig:PID_ks0} (a)) and particles' $m^{2}$ information from Time Of Flight (TOF) (Fig.~\ref{fig:PID_ks0} (b)). A combination of $dE/dx$ and $m^{2}$ criteria is used to identify the particles with purity higher than 96\%.

The $K^{0}_{s}$ particles are reconstructed via a weak decay channel: $K^{0}_{s}$ $\rightarrow$ $\pi^{+}$ + $\pi^{-}$ (Fig.~\ref{fig:PID_ks0}(c)). A set of topological cuts were used to ensure the reconstructed $K^{0}_{s}$ purity higher than 95\%.

\section{Results}
\subsection{$K^{0}_{s}-K^{0}_{s}$ Correlation Functions at $\sqrt{s_{\rm NN}}$ = 39\,GeV and 200\,GeV}

\begin{figure}[htbp]
\begin{center}
    \includegraphics[width=1.0\textwidth]{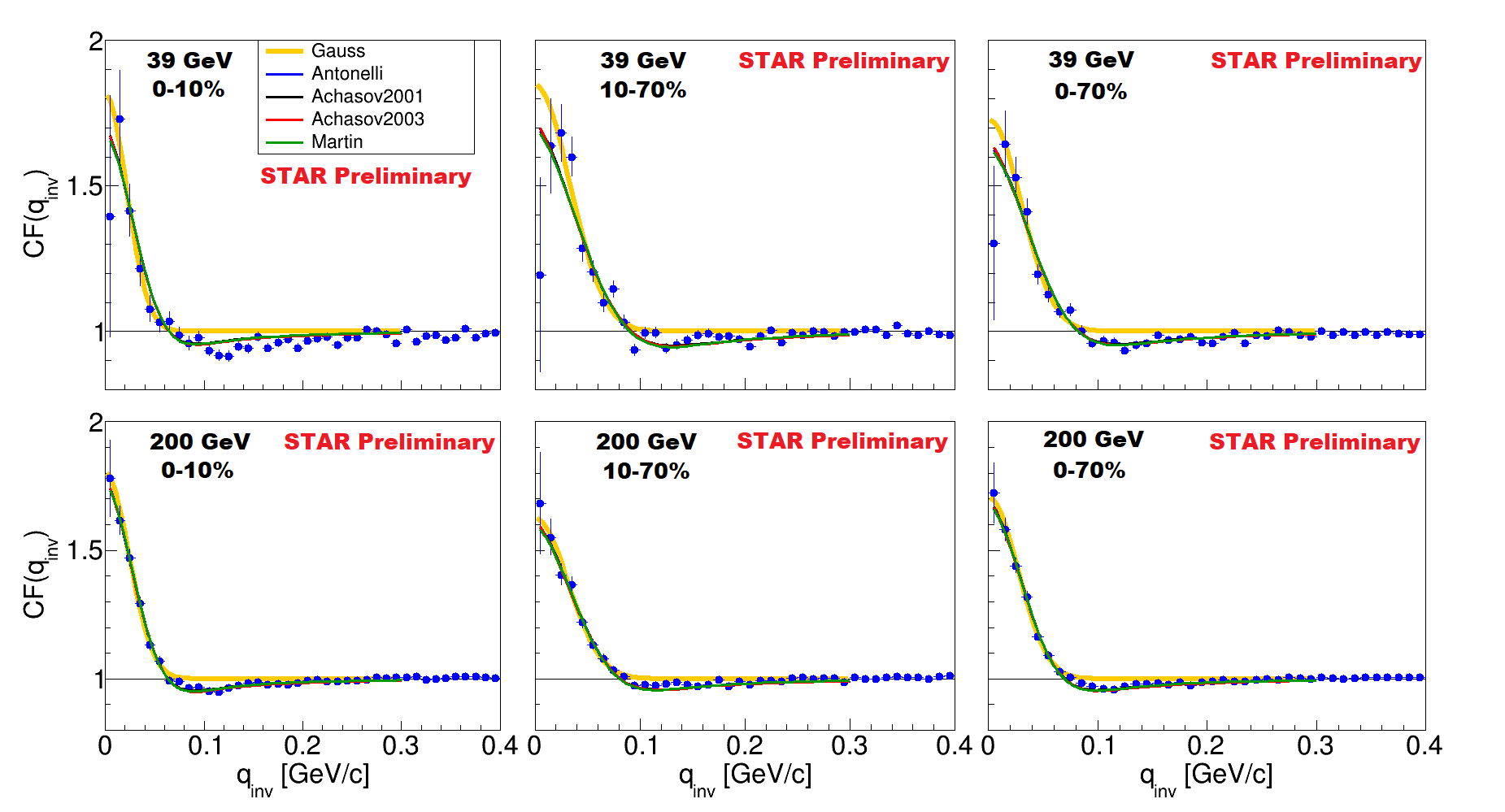}
    \caption{The $K^{0}_{s}-K^{0}_{s}$ correlation functions at $\sqrt{s_{\rm NN}}$ = 39\,GeV (upper row) and 200\,GeV (lower row) with three centrality classes (0-10\%, 10-70\% and 0-70\%) as a function of $q_{inv}$. The statistical errors are shown as vertical lines. The colored lines represent the Gaussian fit and L-L model fit results. }	
	\label{fig:Ks0Ks0_CF_39GeV_200GeV}
	\end{center}
\end{figure}

\begin{figure}[htbp]
\hspace{-0.7cm}
    \includegraphics[width=1.1\textwidth]{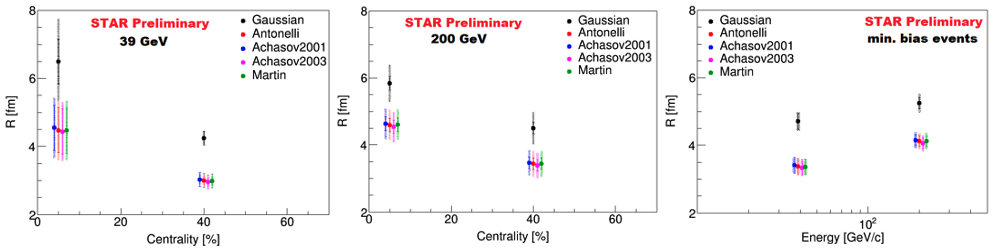}
    \caption{Centrality and energy dependence of $R$ extracted by fitting the Gaussian function and L-L model to $K^{0}_{s}-K^{0}_{s}$ correlation function at $\sqrt{s_{\rm NN}}$ = 39\,GeV and 200\,GeV in Au+Au collisions. The statistics and systematics errors are shown as the vertical lines and shadowed bands, respectively.}
    \label{fig:Ks0Ks0_CF_SourceSize}
\end{figure}

Figure~\ref{fig:Ks0Ks0_CF_39GeV_200GeV} shows the $K^{0}_{s}-K^{0}_{s}$ correlation functions as a function of $q_{inv}$ ($q_{inv}$ = 2$k^{*}$) at $\sqrt{s_{\rm NN}}$ = 39\,GeV and 200\,GeV with three centrality classes (0-10\%, 10-70\%, and 0-70\%) in Au+Au collisions.
The correlation functions show dip structures around $q_{inv}$ = 0.1\,GeV/$c$, which is caused by the near-threshold $f_{0}$(980) and $a_{0}$(980) resonanses. The parameterization is done by using Gaussian function which only includes quantum statistics (QS):
\begin{equation}
C\left(q_{inv}\right)=1+\lambda e^{\left[-R^{2} q_{inv}^{2}\right]},
\end{equation}
and Lednicky-Lyuboshitz model~\cite{Lednicky:2005af} (L-L model) which includes both QS and strong interaction (SI) :
\begin{equation}
\hspace{-1.0cm}
C\left(q_{inv}\right)=1+\lambda\left(e^{\left[-R^{2} q_{inv}^{2}\right]}+\frac{1}{2}\left[\left|\frac{f\left(k^{*}\right)}{R}\right|^{2}+\frac{4 \Re f\left(k^{*}\right)}{\sqrt{\pi} R} F1\left(q_{inv} R\right)-\frac{2 \mathfrak{f} f\left(k^{*}\right)}{\sqrt{\pi} R} F2\left(q_{inv} R\right)\right]\right),
\end{equation}
where $R$ is the invariant radius, $\lambda$ is correlation strength, $f(k^{*})$ is the s-wave $K^{0}\bar{K^{0}}$ scattering amplitude.
In this work, we use the resonance masses and couplings from Ref.~\cite{Martin:1976vx,Baru:2004xg,Antonelli:2002ip,Achasov:2001cj}. The experimental data can be well described by L-L model, while the Gaussian function failed to describe the dip structure. This implies that the strong final state interaction has a significant effect between $K^{0}_{s}-K^{0}_{s}$ pairs.
Figure~\ref{fig:Ks0Ks0_CF_SourceSize} shows the centrality dependence and energy dependence of the extracted radii. The results suggest a decreasing trend from central to peripheral collisions and an increasing trend from low to high energy. 
Also, a significant difference in radii between QS and L-L model is found.


\subsection{Light Nuclei Correlation Functions at $\sqrt{s_{\rm NN}}$ = 3\,GeV}
The production mechanisms of light nuclei are still under debate. Some experimental methods to resolve the problem have been proposed~\cite{Mrowczynski:2020ugu}. Femtoscopy of light nuclei may provide a unique tool to obtain detailed information about the production mechanism.
At low beam energies ($\sqrt{s_{\rm NN}}$ $<$ 20\,GeV), an enhancement in the light nuclei yield is expected due to the high baryon density~\cite{PhysRevC.99.064905}. This provides a great opportunity for precise femtoscopy that involves light nuclei~\cite{Mrowczynski:2021bzy}.

\begin{figure}[htbp]
\hspace{-0.8cm}
    \includegraphics[width=1.1\textwidth]{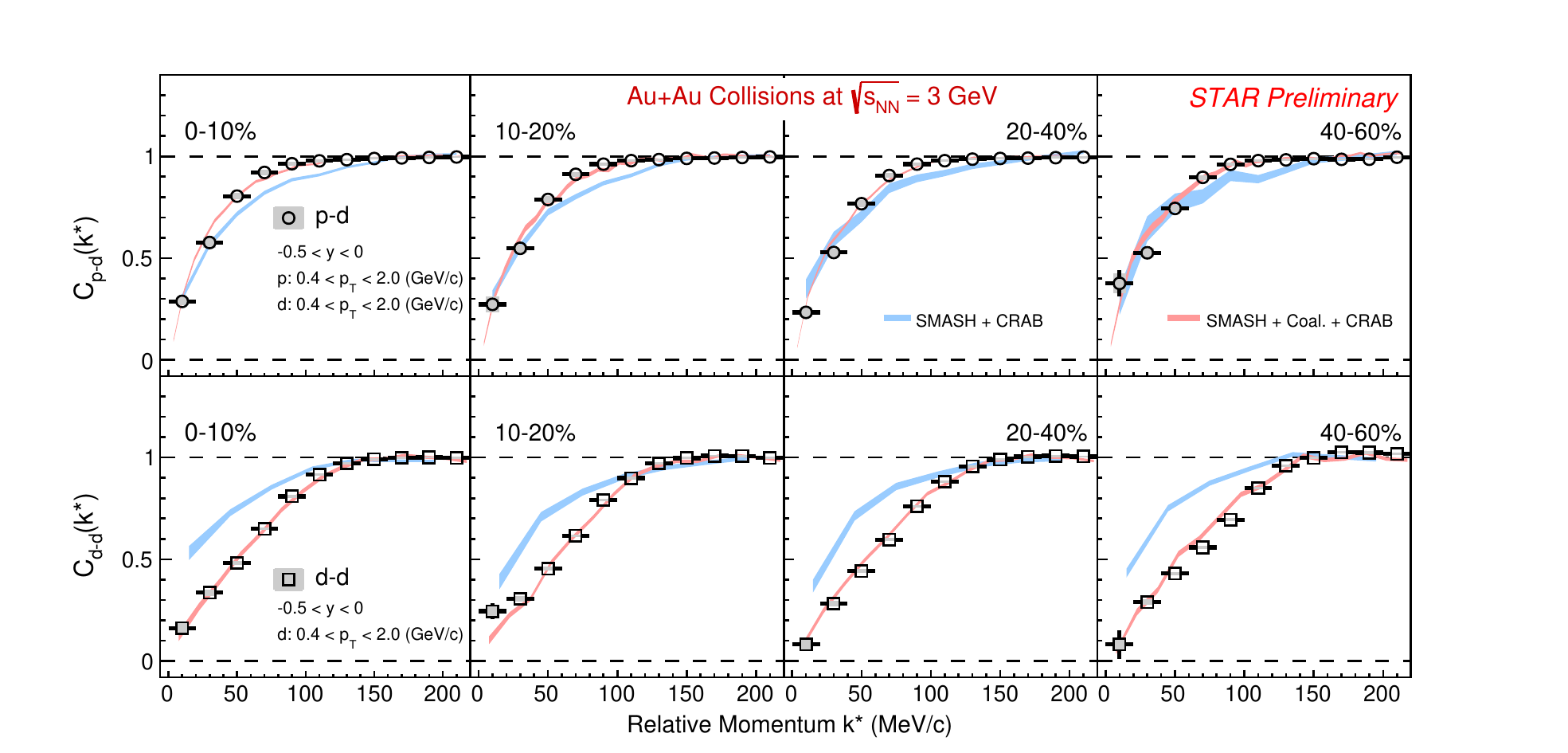}
    \caption{The $p-d$ and $d-d$ correlation functions in different collision centralities in Au+Au collisions at $\sqrt{s_{\rm NN}}$ = 3\,GeV. The statistical and systematic errors are shown as vertical lines and grey bands, respectively. The colored bands represent the $p-d$ and $d-d$ correlations obtained with the deuteron from nucleon coalescence (red) in SMASH and directly produced from SMASH via hadronic scattering (blue), respectively. }
    \label{fig:pd_CF_3GeV}
\end{figure}

Figure~\ref{fig:pd_CF_3GeV} shows the first measurements of $p-d$ and $d-d$ correlation functions at mid-rapidity with four centrality classes in Au+Au collisions at $\sqrt{s_{\rm NN}}$ = 3\,GeV. Clear depletions below unity due to final state interaction at small $k^{*}$ are observed in the measured correlations. Also no significant centrality dependence is found.

To further understand light nucleus formation, a transport model, Simulating Many Accelerated Strongly-interacting Hadrons Model (SMASH)~\cite{Weil:2016zrk}, is used to simulate particle production in $\sqrt{s_{\rm NN}}$ = 3\,GeV Au+Au collisions. We use the cascade mode in SMASH model where the particles are propagated as in vacuum between collisions with other particles. In this work, two different versions of SMASH model are chosen to generate deuterons. In the first version, SMASH can generate deuterons via hadron scattering, such as $p$ + $n$ + $\pi$ $\leftrightarrow$ $d$ + $\pi$. This is so-called directly produced deuterons. The second version of SMASH model produces of nucleons and the deuterons are formed from nucleon coalescence with an afterburner package based on Wigner function~\cite{Zhao:2018lyf}. The SMASH model itself does not contain femtoscopic correlation between two particles after kinetic freeze-out. The interaction (Coulomb potential, strong interaction, and quantum statistics) between two particles is introduced by Correlation Afterburner (CRAB) package~\cite{CRAB}, and the input potentials of $p-d$ and $d-d$ pairs are taken from Ref.~\cite{PhysRevC.33.1303}.

The resulting $p-d$ and $d-d$ correlation functions from the models are shown as color bands in Fig.~\ref{fig:pd_CF_3GeV}, respectively. It is found that the experimental data are well reproduced by the SMASH plus coalescence calculations. On the other hand, the correlation functions with directly produced deuterons from SMASH can only qualitatively reproduce the overall trends, but over/under estimate the data depending on the particle pair. Those observations imply that at $\sqrt{s_{\rm NN}}$ = 3\,GeV Au+Au collisions, deuterons are likely formed via the nucleon coalescence processes. 

\section{Summary}
In summary, we report the femtoscopy results of protons, light nuclei, and strange hadrons in Au+Au collisions from STAR. For $K^{0}_{s}-K^{0}_{s}$ correlations, by fitted with L-L model, it is found that the FSI plays an important role and extracted radii of particle emitting source clear centrality and energy dependence. As for light nuclei correlations, the SMASH model with coalescence can well reproduce data, which suggests that coalescence mechanism may dominate the deuteron production at $\sqrt{s_{\rm NN}}$ = 3\,GeV Au+Au collisions.

\section{Acknowledgments}
 We thank Prof. S. Mrowczynski for insightful discussions about production mechanism of light nuclei. This work was supported by the National Key Research and Development Program of China (2020YFE0202002 and 2018YFE0205201), the National Natural Science Foundation of China (12122505, 11890711) and the Fundamental Research Funds for the Central Universities(CCNU220N003).

\bibliographystyle{ieeetr}
\bibliography{ref}
\end{document}